\documentclass{emulateapj}
\usepackage{bm}
\usepackage{multirow}
\usepackage{color}
\usepackage{natbib}
\usepackage{amsmath}
\usepackage{amsfonts}
\usepackage{amssymb}
\usepackage{epsfig}
\usepackage{hyperref}
\usepackage{graphicx}
\usepackage{booktabs}

\newcommand{\beq}{\begin{equation}}
\newcommand{\eeq}{\end{equation}}
\newcommand{\beqa}{\begin{eqnarray}}
\newcommand{\eeqa}{\end{eqnarray}}

\begin{document}

\title{EXTINCTION AND nebular line properties of A Herschel-selected LENSED DUSTY STARBURST AT $Z=1.027$}
\author{Nicholas Timmons$^{1}$, Asantha Cooray$^1$, Hooshang Nayyeri$^1$, 
Caitlin Casey$^1$, Jae Calanog$^1$, Brian Ma$^1$, Hugo Messias$^2$,Maarten Baes$^3$, R. Shane Bussmann $^{9}$,
Loretta Dunne$^{4,7}$, Simon Dye$^{11}$, Steve Eales$^5$, Hai Fu$^6$, R.J. Ivison$^{7,8}$, Steve Maddox$^{4,7}$, Micha\l{} J. Micha\l{}owski$^{7}$, I. Oteo$^{7,8}$,
Dominik A. Riechers$^{9}$,Elisabetta Valiante$^{5}$, Julie Wardlow$^{10}$ }

\affiliation{$^{1}$Department of Physics  and  Astronomy, University  of California, Irvine, CA 92697}
\affiliation{$^{2}$Centro de Astronomia e Astrof\'ísica da Universidade de Lisboa, O servt\'oŕrio Astr\'omímico de Lisboa, Tapada da Ajuda, 1349-018 Lis- bon, Portugal}
\affiliation{$^{3}$Sterrenkundig Observatorium, UGent, Krijgslaan 281 S9, B-9000 Gent, Belgium}
\affiliation{$^{4}$Department of Physics and Astronomy, University of Canterbury, Private Bag 4800, Christchurch, New Zealand}
\affiliation{$^{5}$School of Physics and Astronomy, Cardiff University, QueensBuildings, The Parade, Cardiff CF24 3AA, UK}
\affiliation{$^{6}$Department of Physics \& Astronomy, University of Iowa, Iowa City, IA 52242}
\affiliation{$^{7}$Institute for Astronomy, Royal Observatory, Blackford Hill, Edinburgh, EH9 3HJ, United Kingdom}
\affiliation{$^{8}$European Southern Observatory, Karl-Schwarzschild-Str. 2, 85748 Garching, Germany}
\affiliation{$^{9}$Department of Astronomy, Cornell University, Ithaca, NY 14853, USA}
\affiliation{$^{10}$Dark Cosmology Centre, Niels Bohr Institute, University of Copenhagen, Denmark}
\affiliation{$^{11}$School of Physics and Astronomy, The University of Nottingham, University Park, Nottingham, NG7 2RD, UK  }

\begin{abstract}

We present Hubble Space Telescope (HST) WFC3 imaging and grism spectroscopy observations of the {\it Herschel}-selected gravitationally-lensed starburst galaxy HATLASJ1429-0028.
The lensing system consists of an edge-on foreground disk galaxy at $z=0.218$ with a nearly complete Einstein ring of
the infrared luminous galaxy at $z=1.027$. The WFC3 spectroscopy with G102 and G141 grisms,
covering the wavelength range of 0.8 to 1.7 $\mu$m, resulted in detections of H$\alpha$+[NII], H$\beta$, [SII], and [OIII] for
the background galaxy from which we measure line fluxes and ratios. The Balmer line ratio H$\alpha$/H$\beta$ of $7.5 \pm 4.4$, when corrected for [NII],
results in an extinction for the starburst galaxy of $E(B-V)=0.8 \pm 0.5$. The $H\alpha$ based star-formation rate, when corrected for extinction,
is $60 \pm 50$ M$_{\sun}$ yr$^{-1}$, lower than the instantaneous 
star-formation rate of 390 $\pm$ 90 M$_{\sun}$ yr$^{-1}$ from the total IR luminosity. 
We also compare the nebular line ratios of HATLASJ1429-0028 with other star-forming and sub-mm bright galaxies. 
The nebular line ratios are consistent with an intrinsic ultra-luminous infrared galaxy with no evidence for excitation by an active galactic nucleus (AGN). 
We estimate the metallicity, $12 + log(O/H)$, of HATLASJ1429-0028 to be 8.49 $\pm$ 0.16. Such a low value is below the average
relations for stellar mass vs. metallicity of galaxies  at $z \sim 1$ for a galaxy with stellar mass of $\sim 2 \times 10^{11}$ M$_{\sun}$.
The combination of high stellar mass, lack of AGN indicators, low metallicity, and the high star-formation rate of HATLASJ1429-0028 suggest that this galaxy 
is currently undergoing a rapid formation.

\keywords
{cosmology: observations --- submillimeter: galaxies --- infrared: galaxies --- galaxies: evolution}
\end{abstract}

\maketitle

\section{Introduction}

Dusty star-bursting galaxies, especially those that are identified at far-IR/sub-mm wavelengths, have infrared luminosities $L_{\rm IR}
\sim 10^{12} - 10^{13}$ L$_{\odot}$, implying star-formation rates (SFRs) in excess of 200 M$_{\odot}$ yr$^{-1}$ (see review
by \citep*{Casey2014}). As a primary contributor to the cosmic far-IR background,
a significant fraction of cosmic star formation and metal production could have occurred in these star-bursting galaxies.
Due to deep and wide surveys with the {\it Herschel} Space Observatory \citep{Pilbratt2010}, we now have 
large samples of dusty, star-burst galaxies at $z > 1$. Despite large number statistics
our knowledge on the physical processes within such galaxies is still limited.

Traditional studies at optical and IR wavelengths involving nebular lines
to probe the interstellar medium (ISM) of these dusty starbursts are challenging due to high dust extinction. One way to overcome this limitation is to
make use of the flux magnification provided by gravitational lensing. Sub-mm surveys provide an efficient way to
select lensed high-redshift galaxies due to the negative K-correction of the thermal dust spectral energy distribution (SED) and
the steep faint-end slope of the sub-mm source counts \citep{Blain1996}. The two large area surveys, {\it Herschel}-ATLAS \citep{Eales2010}
and HerMES \citep{Oliver2012}, have resulted in sufficiently large samples of lensed galaxies \citep{Negrello2010,Wardlow2013,Bussmann2013}
from which we can find interesting targets for detailed follow-up observations. 

Here we present results on the rest-frame optical spectroscopy of a lensed
starburst galaxy to study its nebular line emission and line ratios. We make use of the {\it Hubble} Space Telescope Wide Field Camera 3 (HST/WFC3) grisms for these observations. To detect both H$\alpha$ and H$\beta$ over the wavelength covered by WFC3 grisms we require lensed galaxies to be at $z < 1.6$.
One feasible target for WFC3 grism observations from currently known lensed {\it Herschel} sources is
HATLASJ142935.3-002836 \citep{Messias2014} (H1429-0028; G15v2.19 in \citet{Calanog2014}). The lensed galaxy was detected in the GAMA-15 field of
 {\it Herschel}-ATLAS \citep{Eales2010} with S$_{160\, \mu{\rm m}}$=$1.1 \pm 0.1 Jy$. 
The lensing models of the system using KeckII/NIRC2 laser guide star adaptive
optics image and high-resolution ALMA  imaging data are presented in \citet{Calanog2014}  and \citet{Messias2014}.
The system includes a foreground edge-on disk galaxy (z = 0.218) with a near complete Einstein ring (Fig.~1).
The lens model in \cite{Messias2014} shows that H1429-0028 is comprised of two components with a mass ratio of ($1:2.8^{+1.8}_{-1.5}$).
The two components have been used to suggest H1429-0028 may be undergoing a galaxy merger, but the two components of H1429-0028
are found to lie on top of each other. This also leaves the possibility that the compact bright component of H1429-0028 is a
starbursting clump or a region within a galaxy. The full extent of the galaxy is traced by the extended component that is gravitationally
lensed to an  Einstein ring (Fig.~1). The background source is at z = 1.027 with a total magnification factor of $\mu=7.9 \pm 0.8$  at infrared wavelengths
\citep{Messias2014}. The K-band (AB) magnitude of H1429-0028 is 18.2  \citep{Calanog2014} and is at the level that allows  grism observations with
just one or two HST orbits.

Here we report HST/WFC3 grism spectroscopic observations of H1429-0028 making use of G102 and G141 grism filters,
covering the wavelength range of 0.8 to 1.7 $\mu$m. At the redshift of H1429-0028 these observations probe the rest-frame
wavelength range of 0.4-0.8 $\mu$m allowing us to measure several photoionization emission lines. We use these measurements
to explore the properties of this system in terms of several emission line diagnostic diagrams. 
We also establish the gas-phase metallicity in a star-forming galaxy. The {\it Paper} is organized as following:
In Section~2 we describe the observations and our data reduction procedure. In Section~3 we present
our results related to emission lines and emission line flux ratios and discuss them in the context of
existing studies in the literature. We conclude with a summary in Section~4.
When calculating luminosities we make use of the
standard flat-$\Lambda$CDM cosmological model with H$_0$= 70 km s$^{-1}$ Mpc$^{-1}$ and $\Omega_{\Lambda}$=0.73.

\section{Observations}

\begin{figure}[t]
    \includegraphics[trim = 10mm 10mm 5mm 10mm,clip,scale=0.8]{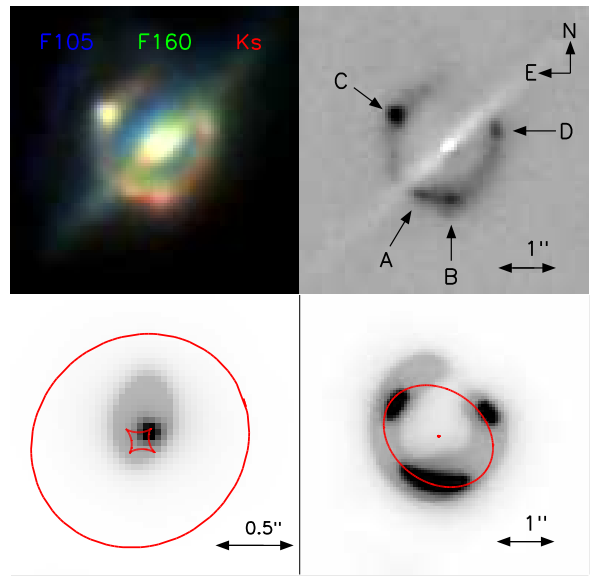}
    \caption{{\it Top Left:} The three color image of the gravitationally lensed system HATLASJ1429-0028 (also G15v2.19 in \citet{Calanog2014})
using WFC3/F105W (blue), F160W (green), and KeckII/NIRC2-LGS K$_s$ (red) imaging data.
 {\it Top Right:} The foreground lensing galaxy was modeled using {\sc GALFIT} \citep{Peng2002} and then 
subtracted from the K$_s$-band imaging data. We label the bright knots following the scheme that was presented in \citet{Messias2014}.
{\it Bottom Left:} Source plane reconstruction showing the two components of HATLASJ1429-0028 and the caustic curves.
{\it Bottom Right:} Image plane reconstruction showing the lens model and the critical curve. Note that the bright features
A, B, C and D are from the bright compact source near the inner cusp caustic while the diffuse ring is due to the extended
source.}
\end{figure}

\begin{figure}[t]
    \includegraphics[scale=0.5]{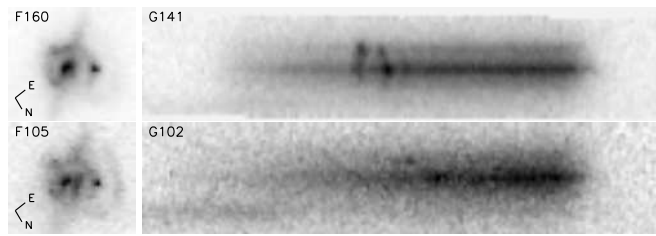}
    \caption{{\it Left:} The direct image in each of the WFC3 imaging filters oriented along the dispersion direction of the grism.
{\it Right:}  The two-dimensional grism images of H1429-0028. The top panel shows G141 and F160W images while the bottom panel
shows G102 and F105W images. The upper trace contains the signal from knots A+B while the lower trace contains the signal from knot C as well as from the foreground lens.}
\end{figure}

HST/WFC3 observations of H1429-0028 were completed with two orbits under GO program 13399 in Cycle 21 (PI:Cooray).
We obtained a total of five exposures, including two direct images and three grism observations in two filters.
The two direct images made use of WFC3/F160W and F105W filters for a total of 250 and 350 seconds, respectively.
We obtained G102 and G141 grism observations over 1800 and 2900 seconds, respectively.
The G141 grism covers 1.0 to 1.8 $\mu$m, while G102 grism covers 0.7 to 1.2 $\mu$m. At $z = 1.027$ these observations then
allow important emission line studies of H1429-0028 involving H$\alpha$ at 1.33 $\mu$m, H$\beta$ at 0.985 $\mu$m,
[OIII] at 1.015 \& 1.005 $\mu$m, and [SII] at 1.364 \& 1.361 $\mu$m. Due to the low spectral resolution of order 80 {\AA} the data do not resolve the [SII] doublet or 
H$\alpha$ from [NII].

We made use of the calibrated HST imaging and grism data from the {\sc CALWF3} reduction pipeline,
as provided by the Space Telescope Science Institute. The spectra for individual objects in the image were extracted with the
{\sc aXe} software package \citep{Kummel2009}.  Briefly, we created an
object catalog making use of the broad-band F160W and F105W images with the {\sc SExtractor}
package \citep{Bertin1996}. A sky background subtraction was performed on 
the direct as well as the grism images. The core {\sc aXe} marks spectral regions for each 
object in the SExtractor catalog, estimates contamination from nearby sources, and flat 
fields each of those regions or beams. A two-dimensional stamp of each grism beam is generated and then combined
together with multiple observations of the same stamp  to create a final two-dimensional image for scientific analysis.
The data products  include the two-dimensional combined grism stamp for each object as well as flux-calibrated 
one-dimensional spectra, contamination estimates, and error estimates. 

We identified emission lines in the one-dimensional spectra  using the known 
redshift of $z = 1.027$ for H1429-0028. We made use of custom IDL scripts as well as the 
Pyraf task {\sc SPLOT} with {\sc ICFIT} to extract the emission line flux densities
and their errors from the one-dimensional spectra. To account for the contamination from
the foreground lensing galaxy, we also model a continuum and account for the contamination in
our line flux densities. Detailed lens modeling of H1429-0028  in four bands have shown clear evidence for
differential magnification. For the rest-frame optical observations as is the case for our data the appropriate magnification is
$7.9 \pm 0.8$ \citep{Messias2014}. 

\begin{figure}
  \includegraphics[scale=.3]{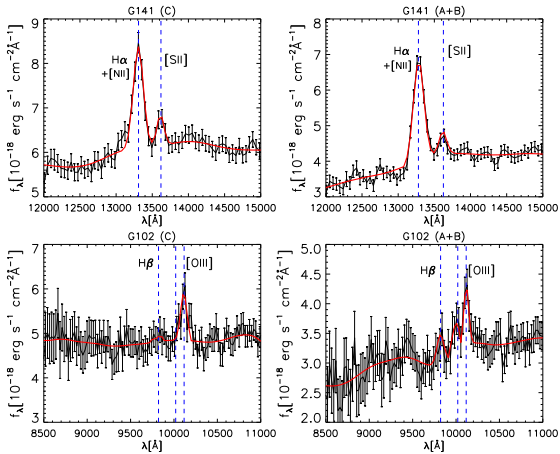}
  \caption {The extracted one-dimensional spectra showing the regions of detected emission lines We separate these detections to different knots
identified in Fig.~1, mainly knot C, brightest of the features, and the combination of knots A+B. 
The spectra from knot C are plotted on the left while the spectra from knots A+B are plotted on the right. Emission lines from knot D
were undetected or confused with the continuum emission from the foreground lensing galaxy in the dispersion direction of the two
grism observations.}
\end{figure}

\section{Results}

\begin{table}[ht]
\caption{Emission Lines} % title of Table
\begin{center} % used for centering table
\begin{tabular}{c c c c} % centered columns (4 columns)
\hline\hline\\ [0.1ex] %inserts double horizontal lines
Line & Component & Flux$^1$ & Eq. Width$^2$ ($\AA$) \\ [0.5ex] % inserts table 
%heading
\hline % inserts single horizontal line
H$\alpha$+[NII] & (A+B) & 45.8 $\pm$ 2.3 & 108.4 $\pm$ 5.4 \\
H$\alpha$+[NII] & (C) & 32.9 $\pm$ 2.6 & 52.4 $\pm$ 4.2 \\ % inserting body of the table
H$\beta$ & (A+B) & 4.5 $\pm$ 2.6 & 15.1 $\pm$ 8.8 \\
H$\beta$ & (C) & 1.4 $\pm$ 3.8 & 2.9 $\pm$ 8.0 \\
$$[OIII]$\lambda$(5007) & (A+B) & 6.4 $\pm$ 1.8 & 18.8 $\pm$ 5.2 \\
$$[OIII]$\lambda$(4959) & (A+B) & 4.1 $\pm$ 1.9 & 12.6 $\pm$ 5.9 \\
$$[OIII]$\lambda$(5007+4959) & (C) & 10.8 $\pm$ 2.2 & 22.6 $\pm$ 4.7 \\
$$[SII](doublet) & (A+B) & 10.1 $\pm$ 1.8 & 23.7 $\pm$ 4.2 \\
$$[SII](doublet) & (C) & 8.7 $\pm$ 2.1 & 14.1 $\pm$ 3.4 \\[1ex] % [1ex] adds vertical space
\hline %inserts single line
\end{tabular}
\end{center}
$^1$ Line fluxes are in $10^{-17}$ erg s$^{-1}$ cm$^{-2}$, not corrected for lens magnification. \\
$^2$ Equivalent widths should be considered as an upper limit due to potential systematic uncertainties in the
background continuum model.\\
\label{table:emission} % is used to refer this table in the text
\end{table}

\begin{figure*}
\begin{center}
  \includegraphics[scale=.65]{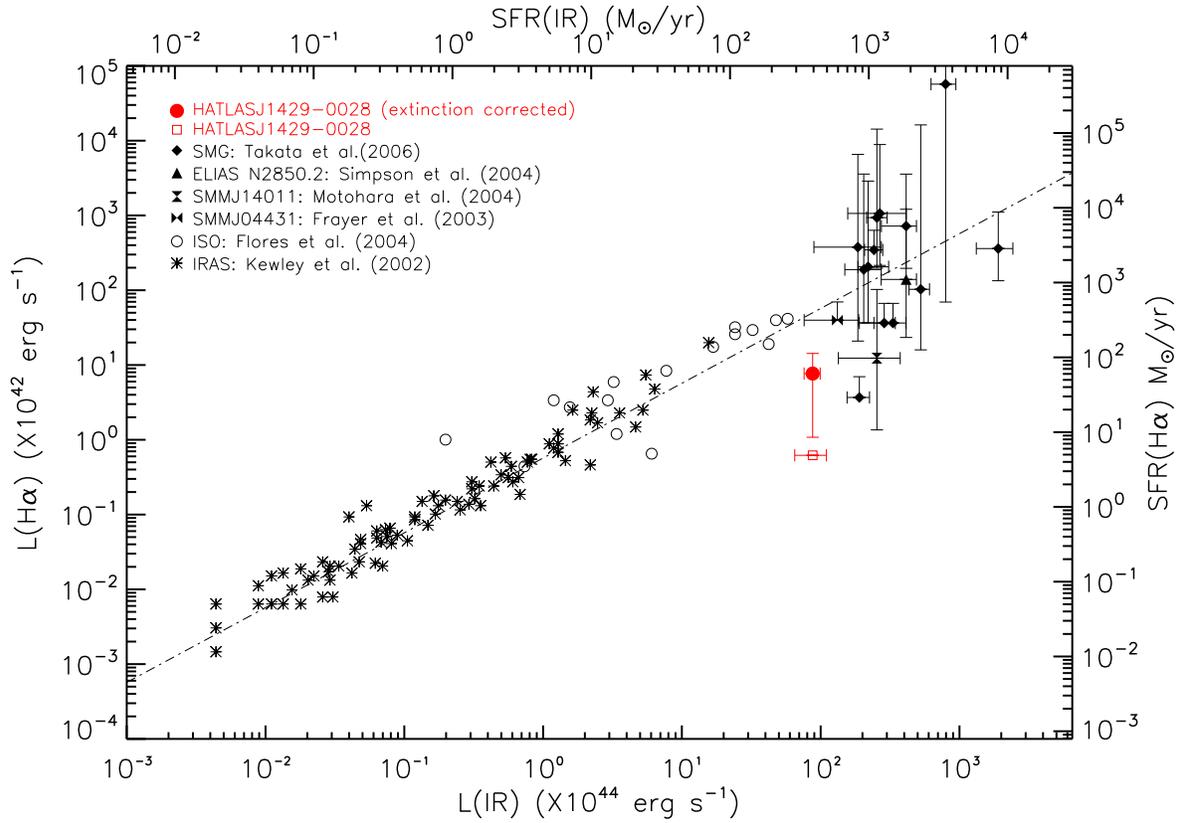}
\end{center}
  \caption {
H$\alpha$ luminosity vs. Far-infrared for H1429-0028 compared to the extinction corrected sample of IR luminous and sub-mm galaxies in \citet{Takata2006}. For reference we also show the corresponding SFRs based on IR luminosity and H$\alpha$ luminosity to the top and left of the plot, respectively. We show H1429-0028  for two cases with and without extinction correction of H$\alpha$ luminosity. The dot-dashed line represents the case that SFRs from H$\alpha$ and far-infrared are equal. H1429-0028  falls below this trend line but the difference between IR and H$\alpha$-based SFRs is fully consistent with the observed scatter of previous measurements.}
\end{figure*}

\begin{figure*}
\begin{center}
  \includegraphics[scale=.9]{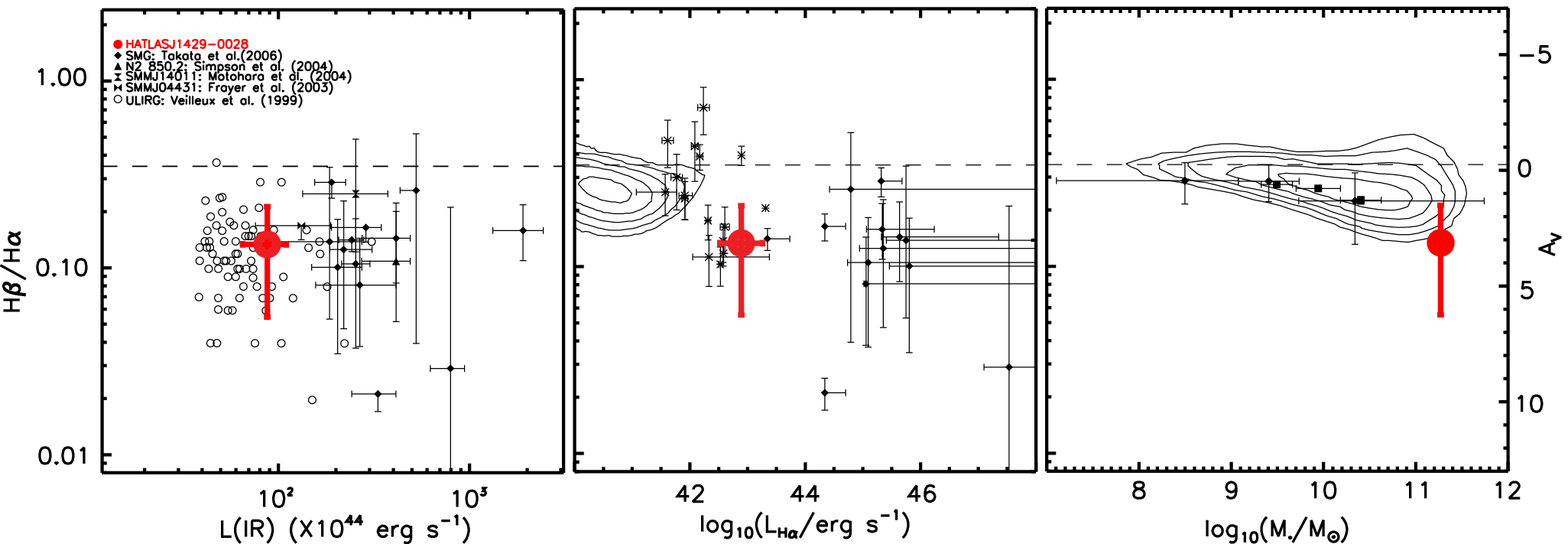}
\end{center}
  \caption {
{\it Left:} Balmer decrement vs. IR luminosity. The background data are from \citet{Takata2006}.
{\it Middle:} Balmer decrement vs. H$\alpha$ luminosity. 
The stars come from  \citet{Dominguez2012} and the diamonds are from \citet{Takata2006}. 
The contours show the galaxy population traced by SDSS.
{\it Right:} Balmer decrement vs. galaxy stellar mass. 
The contours show the galaxy population traced by SDSS. 
The diamonds correspond to  star-forming galaxies of $0.75 \leq z \leq 1.5$ presented in \citet{Dominguez2012}, while the squares
correspond to $z \sim 2$ from \citet{Sobral2012}. We show the expected optical attenuation A$_V$ to the right of the right panel for corresponding values of H$_\beta$/H$_\alpha$.The dashed line represents the intrinsic value of the Balmer decrement.}
\end{figure*}

\begin{figure}
\begin{center}
  \includegraphics[scale=.52]{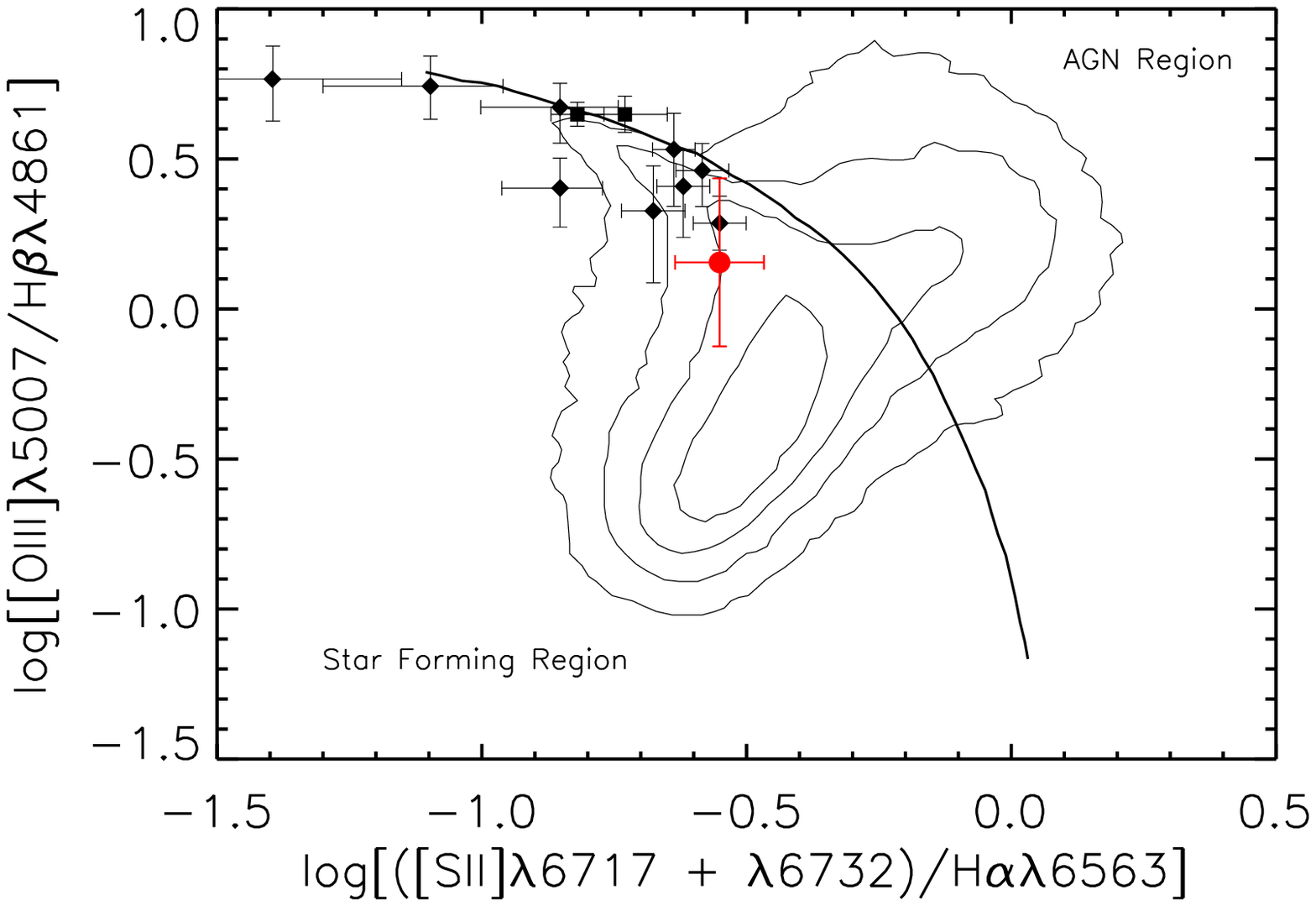}
\end{center}
  \caption {{\it Top:} BPT diagram with the black line separating the AGN and star-forming regions \citet{Kewley2001}.
 The red point corresponds to H1429-0028, while the black points come from star forming galaxies in the redshift range
 $0.75 \leq z \leq 1.5$ from \citet{Dominguez2012} represented as diamonds and two $z \sim 2$ lensed star-forming galaxies from
\citet{Hainline2009} represented as squares. The background contours show the galaxy population traced by
SDSS \citet{Kewley2001} .
}
\end{figure}

\begin{figure}
\begin{center}
  \includegraphics[scale=.52]{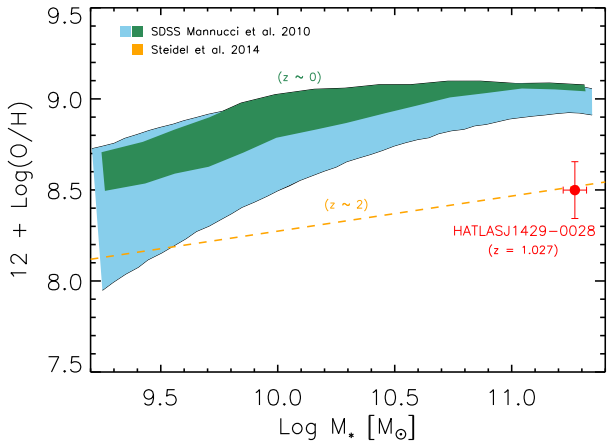}
\end{center}
  \caption { Metallicity vs. Stellar Mass. The green band represents local SDSS galaxies while the blue region
represents a second order fit to SDSS extrapolated towards higher SFR \citep{Mannucci2010}. 
The orange dashed line representing $z \sim 2$ galaxies from \citet{Steidel2014}. The calculated value of metallicity
for H1429-0028 representing the O3N2
calculation using the O[III]/H$\beta$ to N[II]/H$\alpha$ ratio \citep{Pettini2004}.   
}
\end{figure}

\begin{figure}
\begin{center}
  \includegraphics[scale=.52]{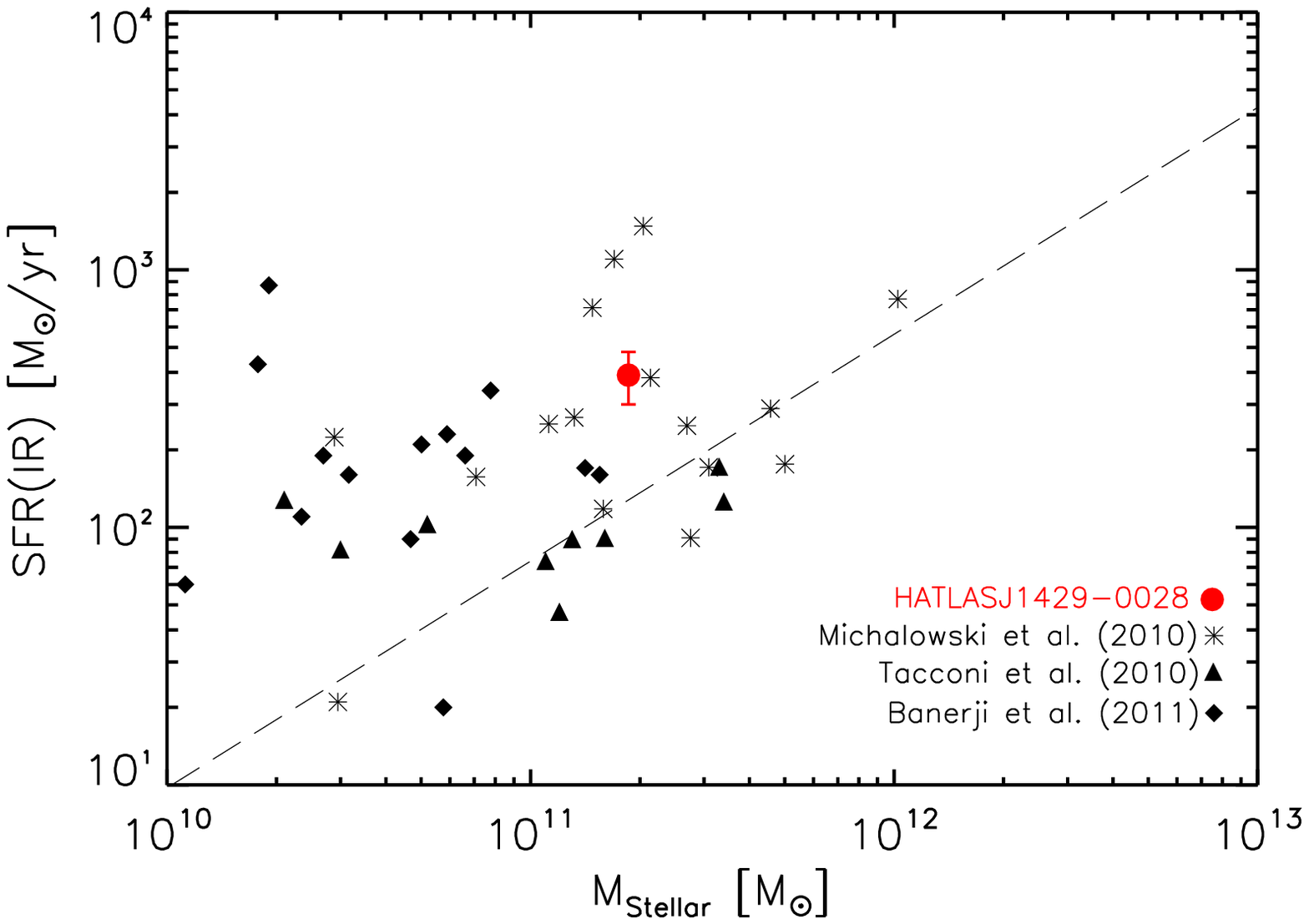}
\end{center}
  \caption {Star formation rate (based on the IR luminosity) vs. stellar mass. For comparison $z \sim 1$ SMGs from \citet{Michalowski2010},\citet{Tacconi2010}, and \citet{Banerji2011} are shown. The dashed line shows the $z=1$ main sequence relation \citep{Elbaz2007}.}
\end{figure}

In order to derive extinction corrections and line ratios, we measure 
the line intensities for detected bright regions of  H1429-0028. In Fig.~1 we label the bright components following the
scheme of \citet{Messias2014}. In the G141 grism the brightest component C and (A+B) as 
well as parts of the Einstein ring had clear emission detections. For G102 only 
components (A+B) had a detectable [OIII] as well as H$\beta$. In G102 component C had clear  detection of the
[OIII] lines but only an upper limit on the H$\beta$ line. 
Given the ratio of H$\alpha$ to H$\beta$ in knots (A+B), and the detected value of H$\alpha$ in knot C, 
the expected value of H$\beta$ in knot C is $\sim 3.2 \pm 1.9 \times 10^{-17}$ erg $s^{-1} cm^{-2}$. 
We measured the H$\beta$ line flux density to be $1.4 \pm 3.8 \times 10^{-17}$ erg $s^{-1} cm^{-2}$. 
This value falls within the estimated range but is not robust enough to be used for scientific analysis.
  
From Fig.~1 the lens models suggest two components for  H1429-0028, one that is compact and bright and a second that is extended.
The two components have effective radii of $0.18" \pm 0.01$ 
and $0.03^{``} \pm 0.01$, for the source responsible for the ring and for knots, respectively. It is clear from the lensing models that the
quadruply imaged knots A to D are from the same source. While it has been suggested that H1429-0028 is a merger, due to the presence of
two components in the lens model, it is not clear from such a model if H1429-0028 is two separate galaxies or
if the smaller component is a high star-forming region within a galaxy. Our spectral line data do not have the adequate velocity resolution
but in the future this question can be addressed with an integral field unit (IFU) observations. In this work, for the line ratios,
we only study the ratios of bright knots. Thus our line ratios capture the physical properties within the starbursting compact region or a compact
galaxy. For the total H$\alpha$ flux  we add the flux from each component with their corresponding magnification from the lens model.
The values are a magnification factor of $\mu$ $\sim 27$ for the compact component, which contributes to the bright knots, 
and $\mu$ $\sim 10$ for the larger component which contributes to the ring. 
We scale the observed line flux to a total estimate of the line intensity across the galaxy
based on K-band photometry of the bright components and the diffuse rings.
This correction results in a factor of $2.6 \pm 0.1$ from the line fluxes measured for the sum of the
components A+B+C to the galaxy as a whole assuming that the continuum fluxes detected for the
other components and ring scale as the rest-frame optical magnitudes. 

To correct for the [NII] contamination of H$\alpha$ we make use of two independent methods to derive the expected
[NII]/H$\alpha$ line ratio and average them as the final value to use here. This follows the approach given in \citet{Dominguez2012}.
The first method from \citet{Sobral2012} estimates the [NII]/H$\alpha$ ratio using the H$\alpha$ + [NII] 
equivalent width (EWs). Using the measured EWs we estimate the ratio to be $0.27 \pm 0.07$.
However it could be that we are overestimating the EWs in our line fitting procedure due to systematic uncertainties
associated with the model for the continuum, especially since the continuum is dominated by the residual
fluxes from the foreground lensing galaxy. Thus we also employ a second method, but we find consistent estimates on
the [NII]/H$\alpha$ ratio.

The second method from \citet{Erb2006} relies on a  relationship derived between stellar mass of a galaxy and
the H$\alpha$/NII ratio of that galaxy. Here, instead of an independent estimate of the stellar mass, we make use of the
SED modeling in Ma et al. (in prep) with a stellar mass of $1.9 \pm 0.02 \times 10^{11}$ M$_{\sun}$ derived from MAGPHYS.
The resulting [NII]/H$\alpha$ ratio is $0.27 \pm 0.03$. Both \cite{Sobral2012} and \cite{Erb2006} make use of the calibration method used in \cite{Pettini2004}.
The two estimates are consistent with each other. Note that the stellar mass we use is slightly lower than the value of $\sim$ $3 \times 10^{11}$ M$_{\sun}$ 
quoted by  \citet{Messias2014}. We prefer to use a revised value for the stellar mass using MAGPHYS models
as we include new optical measurements of the SED and, as discussed later, we find consistent estimates on the extinction with SED modeling
when compared to the estimates based on the Balmer line ratios.

For the values we discuss below we take  the average ratio of [NII]/H$\alpha$ from \citet{Sobral2012} and \citet{Erb2006} to be
$0.27 \pm 0.05$.
Once corrected for [NII] we find H$\alpha$/H$\beta$ = 7.49 $\pm$ 4.4. Using this ratio we calculate the nebular
extinction $E(B-V)$ following  \citet{Momcheva2013} and find it to be $0.82 \pm 0.50$.
Using \citet{Calzetti2001}, the corresponding optical  depth $\tau_V$ is $2.93 \pm 1.9$.
The extinction is lower than the $\tau_V$ value of $\sim 11.2^{+4.5}_{-3.2}$ for H1429-0028 in \citet{Messias2014},
based on the broad-based SED model fitting using {\sc Magphys}. A revised model fit to H1429-0028 
using new estimates of the background galaxy, including deblended IRAC data, finds $\tau_V \sim 4.2 \pm 0.4$ consistent with the
estimate of $\tau_V$ from the Balmer line ratios (Ma et al. in prep). 

In Fig.~4 we compare the extinction-corrected H$\alpha$ luminosity of H1429-0028 with other star-forming galaxies. 
All data points from the literature (following Takata et al. 2006) are extinction corrected though we show the case for
 H1429-0028 with and without extinction correction.
Though the apparent luminosity of  H1429-0028 corresponds to that of a hyper-luminous infrared galaxy with L$_{\rm IR}\sim 10^{13}$ L$_{\sun}$,
the intrinsic luminosity, once corrected for lensing magnification, is that of a ultra-luminous infrared galaxy (ULIRG).
The galaxy falls between the SMGs and local ULIRGs studied by \citet{Swinbank2004} and \citet{Takata2006} with rest-frame optical
spectroscopy at Keck and Subaru, respectively. We find the extinction-corrected SFR of  H1429-0028, at  $60 \pm 50$ M$_{\sun}$ yr$^{-1}$, 
 to be lower than the instantaneous SFR
implied by the total IR luminosity, with a value of 390 $\pm$ 90 M$_{\sun}$ yr$^{-1}$  using the \citet{Kennicutt1998} relation.
 Given the scatter observed in Fig.~4, however, we do not find this difference to be statistically significant.

In Fig.~5 we compare the Balmer decrement of H1429-0028 against H$\alpha$,
IR luminosity, and stellar mass for a sample of galaxies. 
As shown in Fig.~5 middle panel, for the sample of galaxies with both H$\beta$ and H$\alpha$ measurements
in the literature, we find that the extinction-corrected H$\alpha$ luminosity of H1429-0028 to be among the highest.
The \citet{Dominguez2012} sample comes from HST/WFC3 grism observations of $z \sim 0.75$--$1.5$ galaxies. 
The SDSS-detected
star-forming galaxies have  L$_{{\rm H}\alpha} < 10^{42}$ ergs s$^{-1}$, while for H1429-0028  L$_{{\rm H}\alpha} > 10^{43}$ ergs s$^{-1}$.
This is consistent with the fact that  H1429-0028 is an ULIRG. The right panel shows the trend in the Balmer decrement with the
stellar mass such that there is a slight decrease in the   H$\beta$ to H$\alpha$ ratio with an increase in the stellar mass.
The plotted points are the sample-averaged values from \citet{Dominguez2012} as diamonds and
\citet{Sobral2012} as squares in three stellar mass bins in both studies. These data mainly probe the stellar mass below a few times
$10^{10}$ M$_{\sun}$. H1429-0028 is massive with M$_\star \sim 10^{11}$ M$_{\sun}$ and has a Balmer decrement that is lower than the
typical star-forming galaxies in the same redshift range of 0.75 to 1.5.

In Fig.~6 we compare the line ratios of [OIII]/H$\beta$ vs. [SII]/H$\alpha$.
This is a variant of the more traditional BPT diagram \citep{Baldwin1981}
that involves [OIII]/H$\beta$ vs. [NII]/H$\alpha$. Given that [NII] is blended with H$\alpha$ in our low-resolution
data we make use of [SII]/H$\alpha$ ratio. 
We make extinction corrections for the [SII]/H$\alpha$ ratio here given the two [SII] lines are somewhat separated in wavelength from H$\alpha$.
In this diagram  H1429-0028 is consistent with the low metallicity end of the star forming 
regions although the ratios have large uncertainties associated with measurement errors.
The measurements are incompatible with  AGN regions of galaxies from SDSS data at $z < 0.3$.
While H1429-0028 is luminous this is primarily due to gravitational lensing; the intrinsic luminosity of
H1429-0028 is compatible with a galaxy star forming at a rate of 200 to 400 M$_{\sun}$ yr$^{-1}$.
The lens models shown in \citet{Messias2014} are compatible with a merger system. 
Interestingly a value for [NII]/H$\alpha$ of $0.27 \pm 0.03$ is higher than the average [NII]/H$\alpha$ ratio of $0.19 \pm 0.05$
for the galaxies classified as star-forming in the SCUBA sample of \citet{Swinbank2004}, and lower than the average for SMGs hosting
AGN of $0.41 \pm 0.05$ from the same study.

In Fig.~7 we make use of the nebular line ratios, with the estimate of [NII]/H$\alpha$ ratio, to make an estimate of the metallicity.
Instead of an estimate based on [NII]/H$\alpha$ ratio alone, we make use of the O3N2 ratio \citep{Pettini2004} as the estimator here
as it also involves the measured [OIII]/H$\beta$  ratio. The metallicity value, as measured in terms of $12 + \log (O/H)$ 
was found to be $8.49 \pm 0.16$. In Fig.~7 we compare the metallicity vs. the stellar mass. The figure shows the average metallicity vs. stellar mass
relations for both local \citep{Mannucci2010} and $z \sim 2$ galaxies \citep{Steidel2014}. H1429-0028 has 
a metallicity comparable to galaxies at $z \sim 2$ despite being at $z=1.027$.
H1429-0028 has a high SFR, but shows no indication that it is hosting an AGN. H1429-0028 is metal poor despite its high stellar mass and argues for
a scenario that it is still under a rapid formation phase.

Finally in Fig.~8 we show the location of H1429-0028 in comparison to
the main sequence of galaxies at $z \sim 1$. Here we plot  the total IR luminosity-based SFR of
H1429-0028 vs. stellar mass. We find that H1429-0028 is above the $z=1$ correlation from \citet{Elbaz2007}.
For comparison, we also show other dusty star-forming galaxies  at $z \sim 1$ from the literature.  Finally 
the average gas fraction M$_{ISM}$/(M$_{\sun}$+M$_{ISM}$) for the \citet{Tacconi2010} sample of star-forming galaxies 
is 0.34\%. The gas fraction for H1429-0028 is $0.25 \pm 0.1$\%, where we make use of the gas mass of M$_{ISM} = 4.6 \pm 1.7 \times 10^{10}$M$_{\sun}$
from ALMA CO observations reported in \cite{Messias2014}.

H1429-0028 is one example of a grism observation with HST based on a galaxy that was first selected with the
{\it Herschel} catalog as a lensed background source. Based on lensing models \citep{Wardlow2013}, we find that there should be
roughly 0.25 deg$^{-2}$ lensed starburst galaxies in the redshift interval of 1 to 2. In the future such galaxies
will be automatically included as part of the surveys that will be done with slitless grisms on Euclid and WFIRST.
In the 2000 deg$^2$ High Latitude Deep survey we expect WFIRST will detect close to 500 lensed starbursts
at $z \sim 1$ to 3. The study we have presented for one lensed galaxy can then be expanded to a large enough sample for
detailed statistical study that probes the internal structure of lensed starbursts.

\section{Summary}

We observed the {\it Herschel}-selected gravitationally-lensed starburst galaxy HATLASJ1429-0028, studied in detail in \citet{Messias2014} with
some initial description in \citet{Calanog2014}.  We present Hubble/WFC3 G101 and G412 grisms of HATLASJ1429-0028. The observations covered the wavelength regime
of 0.8 to 1.7 $\mu$m and resulted in detections of H$\alpha$+[NII], H$\beta$, [SII], and [OIII] for several bright regions of
the background galaxy. The Balmer line ratio $H\alpha/H\beta$ of $7.5 \pm 4.4$, when corrected for [NII],
results in an extinction for the starburst galaxy of $E(B-V)=0.8 \pm 0.5$. The $H\alpha$ based star formation rate, when corrected for extinction,
is at the level of $60 \pm 50$ M$_{\sun}$ yr$^{-1}$, lower than the star formation rate of 390 $\pm$ 90 M$_{\sun}$ yr$^{-1}$ from the total IR luminosity.
HATLASJ1429-0028 also has a low metallicity despite its high stellar mass at the level of $10^{11}$ M$_{\sun}$.
The combination of high stellar mass, lack of AGN indicators, low metallicity, and the high star-formation rate of  HATLASJ1429-0028 suggests that this galaxy
is still going through a rapid formation.

\acknowledgments

Financial support for this work was provided by NASA
through grant HST-GO-13399 from the Space Telescope Science Institute, which is operated
by Associated Universities for Research in Astronomy, Inc., under NASA contract NAS 5-26555.
Additional support for NT, AC, HN, BM, and CMC was from NSF with AST-1313319.
The Dark Cosmology Centre is funded by the Danish National Research Foundation.
LD, RJI and SJM acknowledge support from the European Research Council Advanced grant COSMICISM.

\bibliography{G15paper-v0.6}

\end{document}